\documentclass[a4paper,12pt]{article}
\usepackage{mathrsfs}
\usepackage{graphicx} 
\usepackage{epstopdf}
\usepackage{subfigure}

\begin{document}
	
	\hoffset = -1truecm \voffset = -2truecm \baselineskip = 10 mm
	
	\title{Exploring the possible gluon condensation signature in gamma-ray emission from pulsars
}
	
	\author{Jianhong Ruan, Zechun Zheng and Wei Zhu\footnote{Corresponding author, E-mail:wzhu@phy.ecnu.edu.cn}
		\\
		\normalsize Department of Physics, East China Normal University,
		Shanghai 200241, China \\}
	
	\date{}
	
	\newpage
	
	\maketitle
	\begin{abstract}
	
	The very high energy gamma-ray emissions from
pulsars are usually considered to be dominated by leptonic scenario since the hadronic flux is weak. We point out that the gluon condensation predicted by a nonlinear QCD evolution equation may greatly enhance the cross sections of proton-target interactions and give rise to a characteristic broken power law in the gamma-ray spectra. The result is used to explore the gluon condensation signature in the observed gamma-ray spectra from pulsars.

\end{abstract}

{\bf keywords}:  Gluon condensation: Astroparticle physics: Gamma ray pulsars:

\vskip 1truecm
\section{Introduction}

Gluons are Bosons. Can gluons be stacked in a same quantum state like Bose-Einstein condensate (BEC)?
How to observe this gluon condensation (GC) phenomenon? Gluons, confined in proton/neutron,
are the fundamental components of the Universe. Previous works concerning the gluon distribution in proton based on QCD have shown the possible existence of
GC [1-3]. The correlations among gluons at very high energy (VHE) may excite chaos in proton, which arises
strong shadowing and antishadowing effects, resulting in the squeeze of gluons to a state with critical
momentum. Accordingly, we found a series of likely GC signals in the VHE cosmic ray (gamma-ray,
electron/positron, proton and nuclei) spectra, the sources include supernova remnants (SNRs), active galactic nuclei (AGN) and
gamma ray bursts (GRBs) [4-7]. The present work focuses on the GC effect in gamma-ray spectra from pulsars.

Pulsars as rapidly rotating and magnetized neutron stars, can generate bubbles of relativistic
particles  when their ultra-relativistic
winds interact with the surrounding medium, and the particles can be accelerated by pulsar wind nebulae (PWNe) up to PeV and beyond.
It is widely accepted that the electromagnetic spectrum of pulsars originates from leptons.
Although protons (or ions more generally) may exist in pulsars and further be accelerated,
the hadronic scenario in pulsars is negligible since their fluxes are very weak in the pulsar surroundings [8].
However, the GC effect may
largely increase the $pp$ cross section by several orders of magnitude [3], which may compensate for the weak proton flux and produce the
observed gamma rays. Therefore, we try to look for the GC signature in the
gamma-rays from pulsars.

In section 2, we give a brief introduction of the GC model and present a complete VHE gamma-ray spectrum predicted by this
model in figure 2, which shows the typical power law on both sides of the sharp turning point
at $E_{\pi}^{GC}$. On the right side of this spectrum there is an exponential cutoff from $E>E_{\pi}^{cut}$.

In sections 3, we try to identify the GC-signature in the gamma-ray spectra of pulsars, and some examples are discussed. Due to the complex structure
of the source, the recorded spectrum may come from different emission mechanisms,
we will only focus on the part that related to the GC-spectrum.

Several VHE sources are tagged as unidentified and located near other bright ones.
Although many works have speculated the possible origins of their spectra, the nature of them is still unknown.
The simple GC-spectrum with a few free parameters allows us to infer the GC signals hidden in other mixed
spectra.  We give such examples in section 4. After the analysis of the above examples, the discussions and summary are presented in section 5.

\section{The GC model}

The flux of high energy gamma ray in hadronic processes $p+p\rightarrow \pi^0\rightarrow
2\gamma$ in the laboratory frame reads [4-7]

$$\Phi_{\gamma}(E)=C_{\gamma}\left(\frac{E}{1GeV}\right)^{-\beta_{\gamma}}
\int_{E_{\pi}^{min}}^{E_{\pi}^{max}}dE_{\pi}
\left(\frac{E_p}{1GeV}\right)^{-\beta_p}$$
$$\times N_{\pi}(E_p,E_{\pi})
\frac{d\omega_{\pi-\gamma}(E_{\pi},E)}{dE},
\eqno(2.1)$$ where the spectral index $\beta_{\gamma}$
denotes the propagating loss of gamma-rays near the source. The accelerated protons obey a power law
$N_p\sim E_p^{-\beta_p}$ with the index $\beta_p$.
$C_{\gamma}$ incorporates the kinematic factor with the flux
dimension and percentage of $\pi^0\rightarrow 2\gamma$. The
normalized spectrum for $\pi^0\rightarrow 2\gamma$ is

$$\frac{d\omega_{\pi-\gamma}(E_{\pi},E)}{dE}
=\frac{2}{\beta_{\pi}
	E_{\pi}}H[E;\frac{1}{2}E_{\pi}(1-\beta_{\pi}),
\frac{1}{2}E_{\pi}(1+\beta_{\pi})], \eqno(2.2)$$where $\beta_{\pi}\sim 1$, and $H(x;a,b)=1$ ( $a\leq x\leq b$) or $H(x;a,b)=0$
(otherwise). Usually, the relations among
$N_{\pi}, E_{\pi}$ and $E_p$ are very complicated and determined by using the limited experimental data.
However, we show that the GC effect simplifies the above relations and gives a special energy spectrum.
Since the more gluons, the more secondary pions in the inelastic $pp$ collision, we assume
that a huge number of gluons at the central region due to
the GC effect may create the maximum number $N_{\pi}$ of pions, which take
up all available kinetic energy if we neglect the other secondary
particles. One can get the following characteristic distributions in the GeV-unit

$$\ln N_{\pi}=0.5\ln E_p+a, ~~\ln N_{\pi}=\ln E_{\pi}+b,  \eqno(2.3)$$
$$~~ where~E_{\pi}
\in [E_{\pi}^{GC},E_{\pi}^{max}], $$and

$$a\equiv 0.5\ln (2m_p)-\ln m_{\pi}+\ln K, ~~b\equiv \ln (2m_p)-2\ln m_{\pi}+\ln K, \eqno(2.4)$$ $K\simeq 0.5$ is the inelasticity.

Substituting equations (2.2)-(2.4) into equation (2.1), we directly get the gamma ray spectrum,

$$E^2\Phi^{GC}_{\gamma}(E)=\left\{
\begin{array}{ll}
\frac{2e^bC_{\gamma}}{2\beta_p-1}(E_{\pi}^{GC})^3\left(\frac{E}{E_{\pi}^{GC}}\right)^{-\beta_{\gamma}+2} \\ {\rm ~~~~~~~~~~~~~~~~~~~~~~~~if~}E\leq E_{\pi}^{GC},\\\\
\frac{2e^bC_{\gamma}}{2\beta_p-1}(E_{\pi}^{GC})^3\left(\frac{E}{E_{\pi}^{GC}}\right)^{-\beta_{\gamma}-2\beta_p+3}
\\ {\rm~~~~~~~~~~~~~~~~~~~~~~~~ if~} E>E_{\pi}^{GC}.
\end{array} \right. .\eqno(2.5)$$Surprisingly, it shows a typical broken power law.

Figure 1 is a schematic diagram for the gluon
rapidity distribution with the GC effect at the $pp$ collision, which is taken from [3].
One can find that the GC effect
begins to work if it's peak locates at $y_{max}=\ln(\sqrt{s_{p-p}^{GC}}/\underline{k}_c)$, thus

$$x_c=\frac{\underline{k}_c}{\sqrt{s_{p-p}^{GC}}}e^{-y_{max}}\simeq \frac{\underline{k}_c^2}
{s_{p-p}^{GC}}.\eqno(2.6)$$Applying $\sqrt{s_{p-p}}=\sqrt{2m_pE_p}$ into equation (2.3), one gets

$$E_{\pi}^{GC}=\exp\left(0.5\ln\frac{\underline{k}^2_c}{2m_px_c}-(b-a)\right).\eqno(2.7)$$

On the other hand, the cross section of the $pp$ collision disappears if the peak of gluon distribution moves
to the rapidity center $y=0$, i.e., at

$$x_c=\frac{\underline{k}_c}{\sqrt{s_{p-p}^{max}}}e^{y=0}, \eqno(2.8)$$or at

$$E_{\pi}^{max}=\exp\left(0.5\ln\frac{\underline{k}^2_c}{2m_px_c^2}-(b-a)\right)=
\frac{E_{\pi}^{GC}}{\sqrt{x_c}}$$
$$=e^{b-a}\sqrt{\frac{2m_p}{\underline{k}_c^2}}(E_{\pi}^{GC})^2=
14(E_{\pi}^{GC})^2.\eqno(2.9)$$ However, figure 1 shows that the plateau is shrinking as
energy increasing rather than expanding, since lots of soft gluons enter the interaction range early
due to the condensation, which implying that the suppression of $\Phi_{\gamma}$ occurs before $E_{\pi}^{max}$. Therefore we add a cut-off factor in equation (2.5), i.e.,

$$E^2\Phi^{GC}_{\gamma}(E)=\left\{
\begin{array}{ll}
\frac{2e^bC_{\gamma}}{2\beta_p-1}(E_{\pi}^{GC})^3\left(\frac{E}{E_{\pi}^{GC}}\right)^{-\beta_{\gamma}+2} \\ {\rm ~~~~~~~~~~~~~~~~~~~~~~~~if~}E\leq E_{\pi}^{GC},\\\\
\frac{2e^bC_{\gamma}}{2\beta_p-1}(E_{\pi}^{GC})^3\left(\frac{E}{E_{\pi}^{GC}}\right)^{-\beta_{\gamma}-2\beta_p+3}
\\ {\rm~~~~~~~~~~~~~~~~~~~~~~~~ if~} E_{\pi}^{GC}<E<E_{\pi}^{cut},\\\\
\frac{2e^bC_{\gamma}}{2\beta_p-1}(E_{\pi}^{GC})^3\left(\frac{E}{E_{\pi}^{GC}}\right)^{-\beta_{\gamma}-2\beta_p+3}
\exp\left(-\frac{E}{E_{\pi}^{cut}}+1\right).
\\ {\rm~~~~~~~~~~~~~~~~~~~~~~~~ if~} E\geq E_{\pi}^{cut},
\end{array} \right. \eqno(2.10)$$or

$$\Phi^{GC}_{\gamma}(E)\equiv\left\{
\begin{array}{ll}
\Phi_0\left(\frac{E}{E_{\pi}^{GC}}\right)^{-\Gamma_1} \\ {\rm ~~~~~~~~~~~~~~~~~~~~~~~~if~}E_{\gamma}\leq E_{\pi}^{GC},\\\\
\Phi_0\left(\frac{E}{E_{\pi}^{GC}}\right)^{-\Gamma_2}
\\ {\rm~~~~~~~~~~~~~~~~~~~~~~~~ if~} E_{\pi}^{GC}<E_{\gamma}<E_{\pi}^{cut},\\\\
\Phi_0\left(\frac{E}{E_{\pi}^{GC}}\right)^{-\Gamma_2}
\exp\left(-\frac{E}{E_{\pi}^{cut}}+1\right),\\
\\ {\rm~~~~~~~~~~~~~~~~~~~~~~~~ if~} E\geq E_{\pi}^{cut},
\end{array} \right. \eqno(2.11)$$
where

$$E_{\pi}^{cut}=\alpha E_{\pi}^{max}, \eqno(2.12)$$ we assume $\alpha=0.1$ in this work.
Note that the energy is the GeV-unit. A schematic GC-spectrum is presented in figure 2.

Although the breaks in spectrum with an exponential cutoff are ubiquitous either for the leptonic or hadronic scenarios,
the GC-spectrum (2.11) has its own features. (i) The shape of gamma-ray spectra in leptonic and traditional (without the GC effect) hadronic models sensitively depend
on the distribution of parent particles (electrons or protons), while the GC-spectrum is characterized by a few free
parameters ($\Phi_0, \Gamma_1, \Gamma_2, \alpha$ and $E_{\pi}^{GC}$), where the incalculable $pp$ interactions are
simplified by the saturation condition in equation (2.3). (ii) On both sides of the sharp breaking
point $E_{\pi}^{GC}$ the spectrum appears as the single power-law.
(iii) The GC-threshold $E_{\pi}^{GC}$ in equation (2.10) is nuclear target-dependent, which may be from $100~GeV$
to $TeV$ and beyond (see discussions in section 5). (iv) Equation (2.10)
is an analytical solution of the GC model, where the parameters have their
physical meaning and we will use them to predict the undetected
spectra in section 4.

\section{HESSs  J1825-137, J1640-465, J1809-193, J1813-178 and PSR J0205+6449/3C 58}

Although a lot of data about the VHE  gamma ray spectra of pulsars have been reported, however, 
leptonic emission scenarios are usually favored
over hadronic emission scenarios.
Besides, astronomical radiations are complex and
different emission mechanisms may exist in a same source simultaneously. We found the following five examples in the existing data showing a complete
power-law as shown in figure 2.

HESS J1825-137 is one of the most efficient TeV gamma ray emitting PWNes [9].
It was detected by the Fermi Large Area Telescope(Fermi-LAT) in the GeV energy band,
and usually discussed in connection with leptonic acceleration scenarios [10] since it is not a prime candidate for pion decay according to the traditional hadronic scenario. We found that HESS J1825-137 with the
Fermi-LAT data in $10 ~GeV-10 ~TeV$ can also be fitted using the
GC-spectrum (see figure 3), where the
$E_{\pi}^{cut}=14~TeV$ is fixed by equation (2.12). For comparison, we present a leptonic spectrum in figure 3 (dashed curve).

HESS J1640-465 is one of the TeV sources discovered by the HESS survey in inner Galaxy [11]. It's spectrum, explained in the traditional hadronic model [12,13],
is connected to a hard Fermi-LAT gamma-ray spectrum and it forms a long flat spectrum from $100~MeV$ to $1~TeV$. However, Y.L. Xin et al
have suspected that there could not be such a strong proton flux in pulsars [14]. They
reanalyzed the relating Fermi-LAT data and found that an extended GeV gamma-ray source was coincident with J1640-465. Its photon spectrum was described by a power-law with an index of 1.42$\pm$0.19 in the energy range of $10-500~GeV$, and smoothly connected
with the TeV spectrum of HESS J1640-465.  Using the same Fermi-LAT data  we describe the expanded spectrum of HESS J1640-465 in the GC model, and the result is presented in figure 4 (solid curve).

HESS J1809-193 was initially discovered during the systematic search for VHE emission from pulsars in
the Galactic Plane Survey performed with HESS [15].
The environment of HESS J1809-193 is complicated, containing several different sources.
Combining with the Fermi LAT data, the spectrum presents a broad energy plateau in $300~GeV-30~TeV$, the nature of which with such a shape has been regarded as the traditional hadronic model but with a harder proton index. We take an alternate way, i.e,
assuming this broadly extended gamma-ray spectrum might be produced by the leptonic radiation and the hadronic model
with the GC effect. To begin with, we use the precise data for $dN/dE\sim E^{-\Gamma_2}$ [8] and equation (2.11) to obtain $\Gamma_2=2.22$ (see figure 5). Figure 6 is our result using the GC spectrum (2.11), the data are taken from [14].

A high energy pulsar PSR J1813-178 is classified as a PWN candidate, which has a similar spectrum of figure 2.
A new
detection of extended emission in the
range $0.5-500~ GeV$ finds that it is difficult to explain the spectrum
with inverse Compton (IC) emission from high-energy electrons alone [16].
Our predictions of GC (mixing with leptonic models)
are presented in figure 7.

3C 58, classified as a pulsar wind nebula [17], is a complex radio source with an extended flat spectrum.
PSR J0205+6449 is a pulsar near 3C 58,
it's flux is divided into two parts: off-peak and on-peak.
We connect the off-peak part of PSR J0205+6449 with the flux of 3C 58, the result can be fitted by the GC model
as shown in figure 8, and the on-peak part can be described by other leptonic model.

\section{HESSs J1826-130, J1641-463 and J1741-302}

There are several VHE sources presenting a typical power law with the hard photon index $\Gamma_{\gamma}\sim 2$
in a broad energy range from hundreds of
GeV to beyond 10 TeV. Besides, all of them are tagged as unidentified and located near other bright VHE sources.
Although many works have speculated the possible origins of these VHE gamma-ray spectra, their nature is still open.
Especially, where and how cosmic rays are accelerated to reach the necessary PeV energies and beyond in our Galaxy, and in which mechanism
the kinetic energy of the initial particles (protons or electrons) converts into photons
are still unknown. An important obstacle is that we lack the extension of
the above spectra in the GeV-energy region, and an incomplete spectrum cannot provide the correct information of these sources.
The GC-model predicts the typical broken power law, it reduces the number of free parameters
and accordingly the uncertainty. We will use the GC-model
to predict the incomplete spectra of the following three examples.

J1826-130 is an unidentified hard spectrum source discovered by HESS along the Galactic plane [18],
and was previously hidden in the extended tail of emission from the nearby bright source HESS J1825-137.
HESS J1826-130 shows a relatively hard gamma-ray spectrum, which obviously incomplete. However we can perfect
it by using the GC model. Our logical reasoning is:
(i) Using equation (2.11) at $E_{\gamma}>E_{\pi}^{GC}$ to fit the HESS J1826-130 data, one can find that
a single power law with $\Gamma_{\gamma}\sim 2$
extends up to $14~TeV$, and then appears a cut-off factor (see figure 9). According to equations (2.9) and (2.12), the
breaking energy is $E_{\pi}^{GC}=100~GeV$.
(ii) We turn to HESS J1825-137 for information related to HESS J1826-130.
Note that the parameter $\beta_p$ in the GC model is determined by the accelerate mechanism of protons. Therefore, we assume that the two sources are close to each other and
have a same value $\beta_p$. Thus, we have
$\Gamma_2=-\beta_{\gamma}+2\times 0.91-1=2 $ for HESS J1826-130. As a result a complete spectrum of
J1826-130 can be shown in figure 10 since we already know $\beta_{\gamma}=1.18$.

Why there is no recorded spectrum of HESS J1826-130 at $E_{\gamma}<100~GeV$? We noticed that a gamma-ray pulsar J1826-1256
was discovered by Fermi-LAT [19], and
it's spectrum at the GeV energies is shown
in figure 10, which may shadow the spectrum of HESS J1826-130 in the observation and leads to an incomplete spectrum.

A similar example is HESS J1641-463 [20], whose spectrum is also unclear
due to the confusion with the bright nearby source HESS J1640-465 [11]. No $X$-ray candidate stands out as a clear association. Usually
it's spectrum is explained by hadronic mechanism, where the emission
is produced by cosmic ray protons colliding with the ambient gas [21].
The data in figure 11 tells us that $\Gamma_2=2.07$, $E_{\pi}^{cut}=14~TeV$
in the GC model, and accordingly $E_{\pi}^{GC}=0.1~TeV$ by equation (2.9). We take $\beta_p=1.02$ in HESS J1641-463 as in HESS J1640-465
assuming their proton flux are originated from a same accelerator, consequently $\beta_{\gamma}=0.99$. Similarly, a complete spectrum of J1641-463 is presented in figure 12.
The hollow points are the detected pulsed emission of SNR G338.5+0.1
by Fermi-LAT [22], we add it in figure 12 for they
may cover the extension of the HESS spectrum in the GeV energies.

A preliminary detection of HESS J1741-302 was early announced by HESS [23]
and thanks to the increased amount of high quality VHE data and improved analysis techniques,
the spectrum of it can now be analyzed in detail [21]. HESS J1741-302 is not only unidentified but also has no plausible counterpart  at the high-energy
($0.1 ~GeV-100~ GeV$) range,
the region around it is rather complex, as a compact radio source and a variable star
are spatially coincident with the position of HESS J1741-302.
There is only one incomplete spectrum as shown in figure 13, again using
the GC model we try to unveil its nature. From the gamma ray  distribution in figure 13,
one can find that $\beta_{\gamma}+2\beta_p-1=\Gamma_2=2.3$, and $E_{\pi}^{cut}\simeq 14~TeV$ corresponds to $E_{\pi}^{GC}=0.1~TeV$. Thus, we have got $\beta_p=0.91$ and $1.02$ for J1826-130 and J1641-463, differed by 0.11. Besides,
the measured value of $\Gamma_2$ for these two sources has an error of $0.11\pm 0.20$.
Therefore, we assume that the above three VHE gamma
rays originate from the same proton accelerate mechanism in our Galaxy and we take $\beta_p\simeq 1$  for J1741-302.
Thus, a complete VHE gamma ray spectrum can be predicted in figure 14.
An investigation of Fermi LAT data has revealed a new high energy source Fermi J1740.1-3013 [24], which is $\sim 0.3^o$ offset from
the best fit position of HESS J1741-302.  We presented it in figure 14 considering that the spectrum of HESS J1741-302 at GeV energies may be
covered or mixed by this new source.

\section{Discussions and Summary}

It is generally believed that the gamma rays in pulsar environment are mainly generated by leptonic mechanism
since the proton flux is weak. However, with the GC model, the
above point of view could be modified. The increment of the $pp$ cross section due to
the GC effect may compensate for the lack of proton flux and emit
gamma-rays with the GC-characteristics. On the other hand, not all the hadronic accelerators in the Galaxy can generate the GC effect, because comparing with the traditional hadronic model, the GC effect needs
higher energy of parent protons to convert into more mesons.
Therefore, the hadronic scenario with the GC effect is not as common as leptonic scenario in pulsars.

The inverse Compton emission (or curvature radiation) accompanied by synchrotron X-ray spectrum is an important
feature of leptonic model for VHE gamma-ray radiation. The GC model has no such correlation but has a
much deeper connection with other processes. In fact, the GC spectrum (2.11) can be used to explain the VHE gamma ray spectra
in supernova remnant (SNR) [5], active galactic nuclei (AGN) [6] and gamma ray burst(GRB) [6]. Besides, adding the process $\gamma
\rightarrow e^-+e^+$ or a hadronization mechanism, the same GC model can investigate anomalous excess in electron-positron spectra [5]
or hadronic fluxes [7].

A question needs to be explained is that different processes may give different parameters in
equation (2.11), which are related to the structure and nature of the source, especially, the important parameter $E_{\pi}^{GC}\simeq
400~GeV$ for Tycho's SNR while $E_{\pi}^{GC}\simeq 100~GeV$ for the sources in this work.
Why these $E_{\pi}^{GC}$ are different is because that
they are target nuclei number $A$-dependent: $E_{\pi}^{GC}(p-A)>E_{\pi}^{GC}(p-A')$ (if $A<A'$)
since the nonlinear corrections enhance as $A$ increases.
A roughly simulation of the GC model shows that
GC-threshold $E_{\pi}^{GC}$ drops rapidly from $A=1$ and then decrease slowly from the intermediate nucleus [3,4].
We
divide the targets into a few of categories and assume that on average, target nuclei near pulsar is
heavier than that in Tycho's SNR. As discussed above, the broken points $E_{\pi}^{GC}$ in the GC model
are concentrated around a few different values.
A detailed estimation requires more precise examples to be collected. The spectrum
of the investigated pulsars is compatible with being produced through the
GC effect, compensating for a weak flux of hadrons.
Therefore, as shown in figures 4, 6-8, 10, 12 and 14, the GC-spectrum is always accompanied by a leptonic spectrum, although the leptonic fit
is beyond the scope of this work. The parameters used/fitted for each source are presented in table 1.
An exception in our examples is HESS J1825-137, it extends to a long distance from the central pulsar PSR J1826-1334, where the electrons may escape from the nebula and diffuse in the interstellar medium [25].

In summary, the gluon condensation originating from a nonlinear QCD evolution equation
may greatly increase the cross sections of
proton-target interactions and arise a characteristic broken power law in the gamma-ray spectra. The result is used to explore the gluon condensation signature in the observed gamma-ray spectra from pulsars.

{\it Acknowledgments:} We thank Zhiyi Cui and Lihong Wan for useful comments.
This work is supported by the National
Natural Science of China (No.11851303). \\\\

\newpage

\begin{figure}
	\begin{center}
		\includegraphics[width=0.8\textwidth]{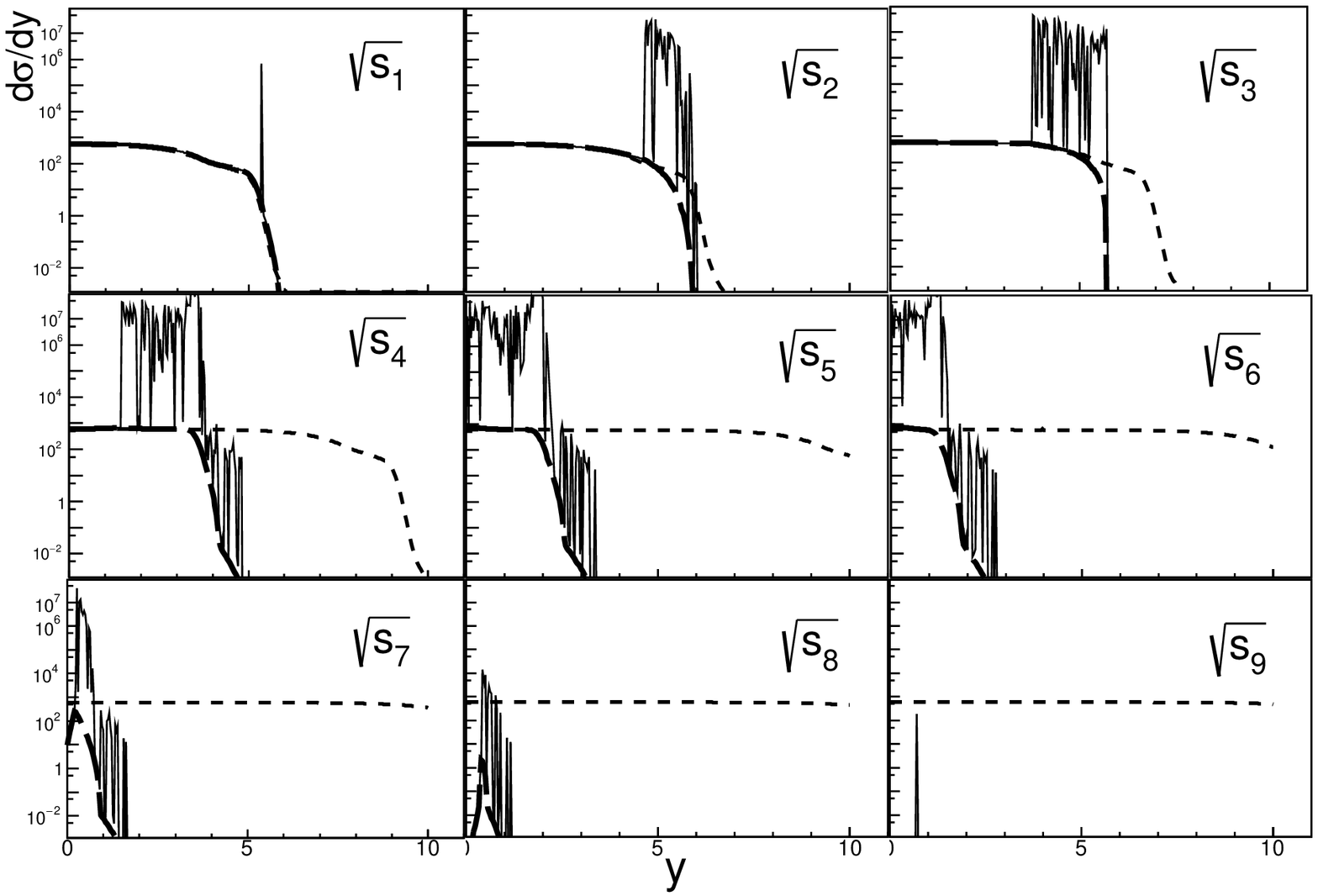} 
		\caption{A schematic diagram for the inclusive gluon
			rapidity distribution at the $pp$ collision [3],
			where $\sqrt{s_{i+1}}>\sqrt{s_i}$. The results show the large
			fluctuations arisen by the GC effect.
		}\label{fig:1}
	\end{center}
\end{figure}

\begin{figure}
	\begin{center}
		\includegraphics[width=0.8\textwidth]{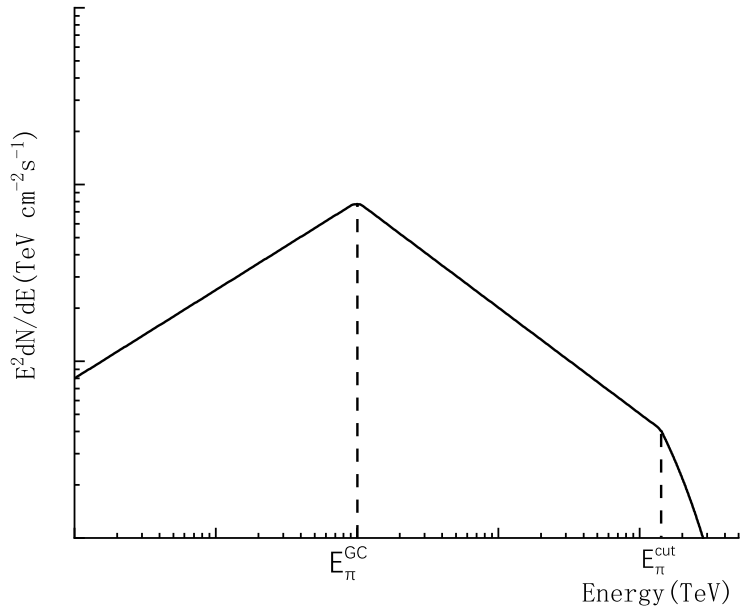}
		\caption{A schematic shape of a complete GC-spectrum according to equation (2.11).
		}\label{fig:2}
	\end{center}
\end{figure}

\begin{figure}
	\begin{center}
		\includegraphics[width=0.8\textwidth]{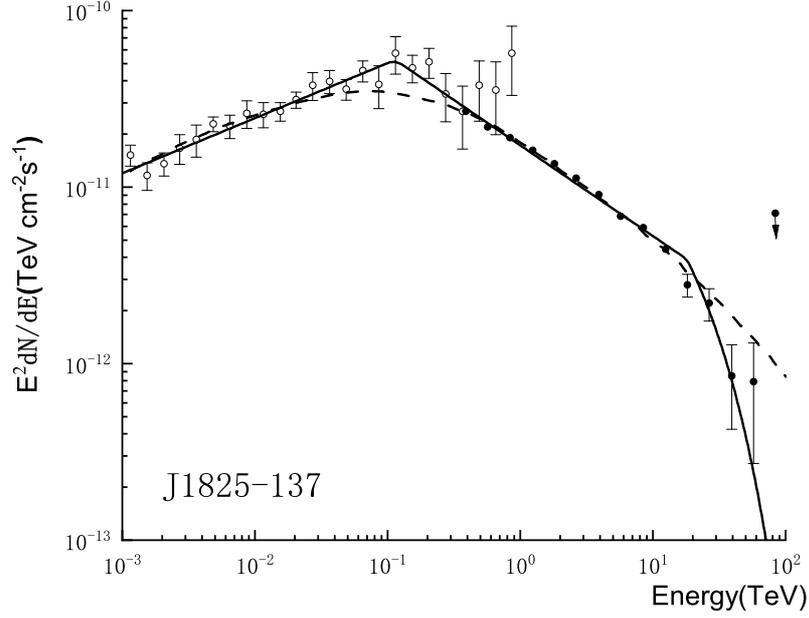} %
		\caption{Gamma-ray spectrum of HESS J1825-137 (black points) combined with the Fermi-LAT data
			(hollow points) [9]. The solid curve is obtained by the GC model,
			and the dashed curve is
			the prediction of a leptonic model [10].
		}\label{fig:3}
	\end{center}
\end{figure}

\begin{figure}
	\begin{center}
		\includegraphics[width=0.8\textwidth]{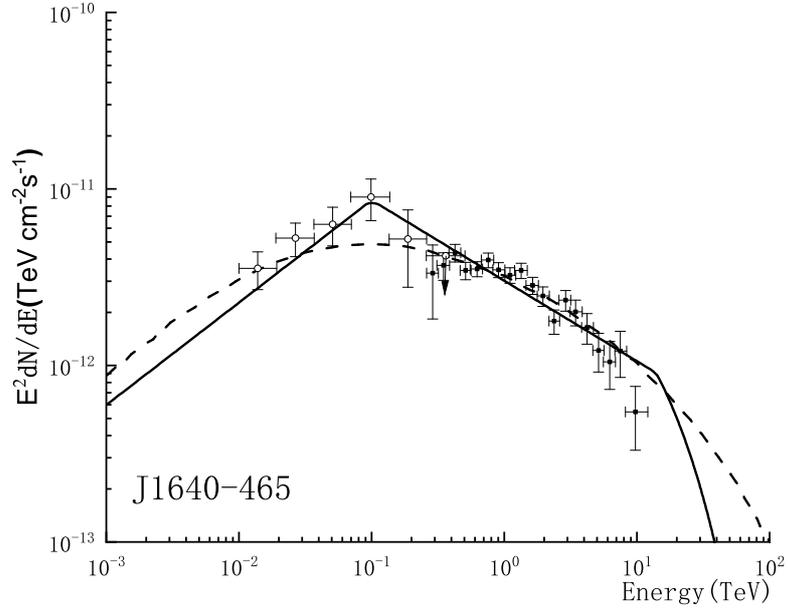} %
		\caption{Gamma-ray  spectrum of HESS J1640-465 (black points) combined with the Fermi-LAT data
			(hollow points) [11]. The solid curve is obtained by the GC model.
			The dashed curve is the prediction of a leptonic model [14].
		}\label{fig:4}
	\end{center}
\end{figure}

\begin{figure}
	\begin{center}
		\includegraphics[width=0.8\textwidth]{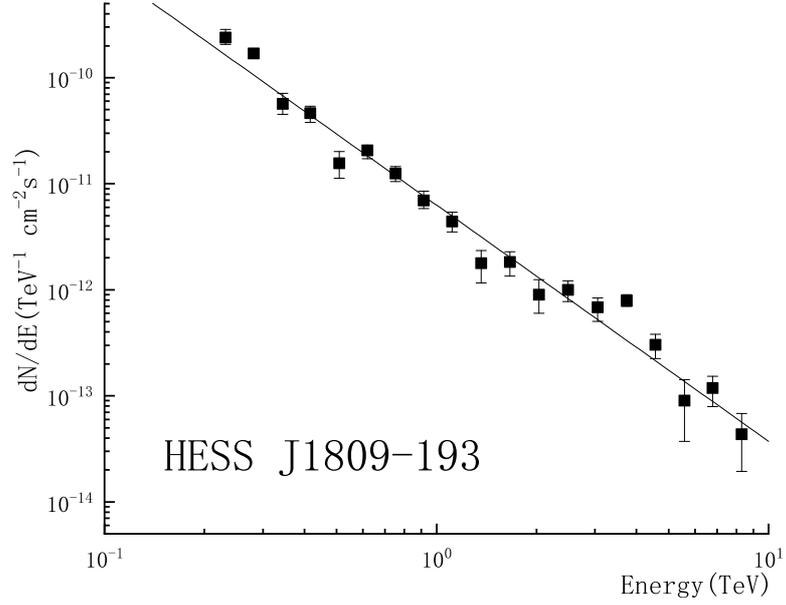} %
		\caption{A part of gamma-ray spectrum of HESS J1809-193 [9], which gives $\Gamma_2=2.22$.
		}\label{fig:5}
	\end{center}
\end{figure}

\begin{figure}
	\begin{center}
		\includegraphics[width=0.8\textwidth]{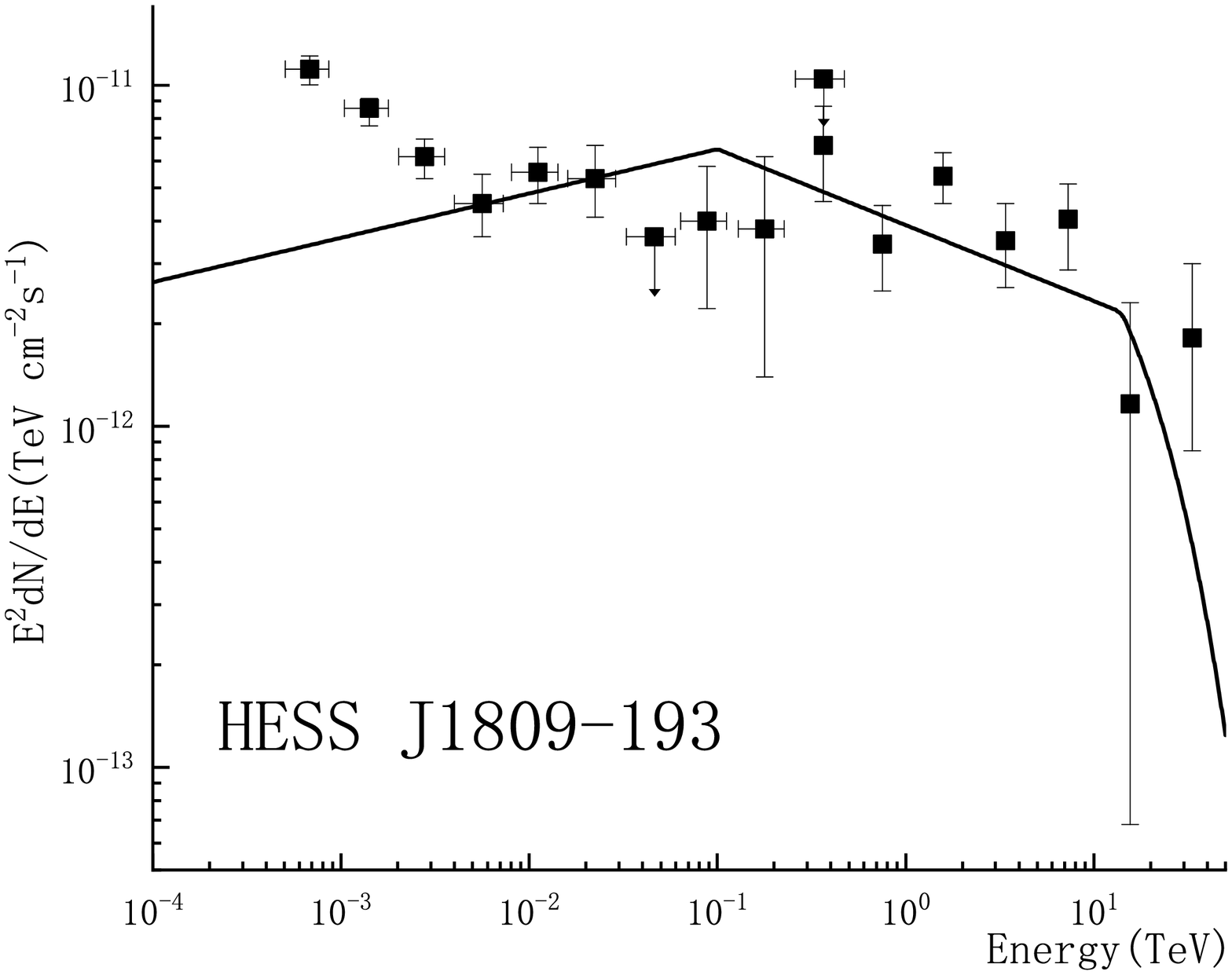} %
		\caption{Gamma-ray  spectrum of HESS J1809-193 [15]. The solid curve is obtained by the GC model.
			The experimental points deviated from the GC-spectrum at $E<10~GeV$
			may originate from other radiation mechanism.
		}\label{fig:6}
	\end{center}
\end{figure}

\begin{figure}
	\begin{center}
		\includegraphics[width=0.8\textwidth]{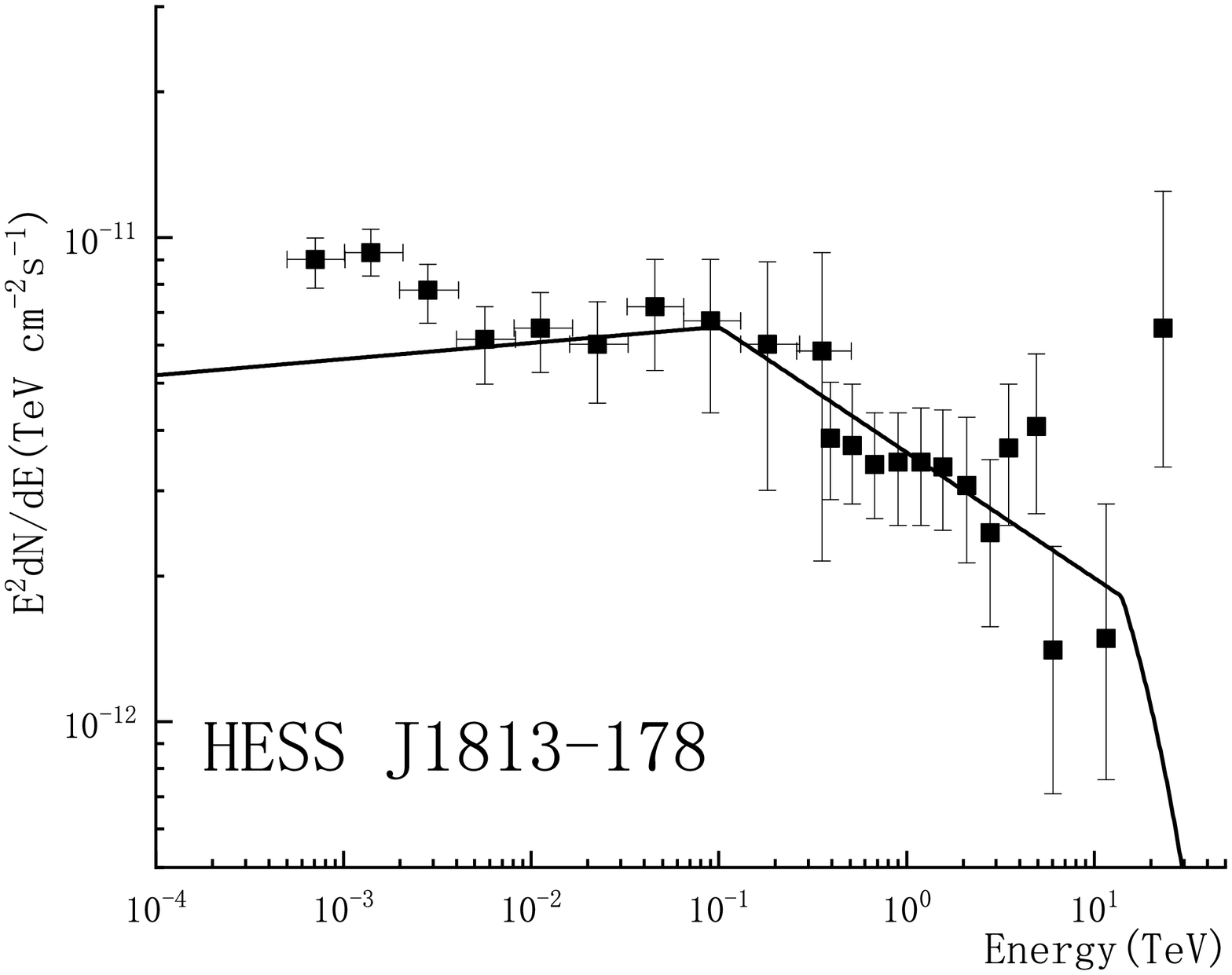} %
		\caption{Gamma-ray spectrum of HESS J1813-178 [16]. The solid curve is obtained by the GC model.
			The experimental points deviated from the GC-spectrum at $E<10~GeV$
			may originate from other radiation mechanism.
		}\label{fig:7}
	\end{center}
\end{figure}

\begin{figure}
	\begin{center}
		\includegraphics[width=0.8\textwidth]{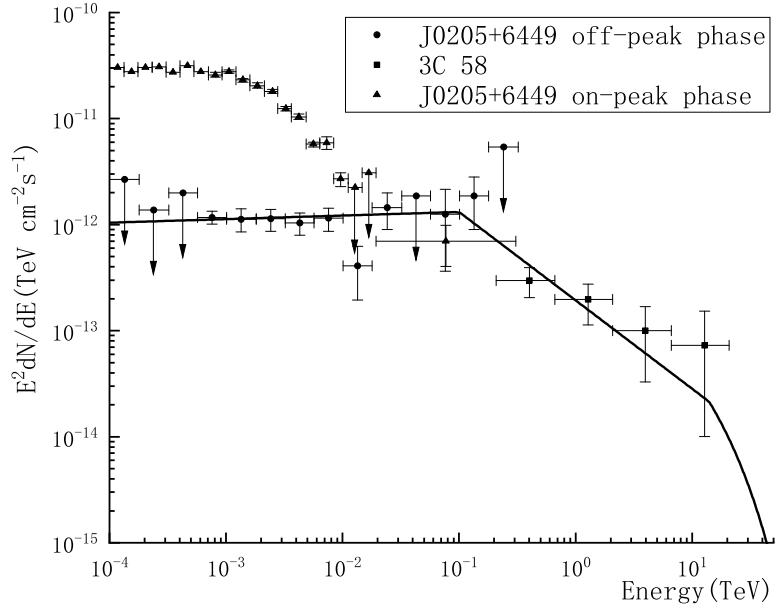} %
		\caption{Gamma-ray distribution of 3C 58 combined with off-peak part of PSR J0205+6449 [17].
			The solid curve is obtained by the GC model. The on-peak points
			may originate from other radiation mechanism.
		}\label{fig:8}
	\end{center}
\end{figure}

\begin{figure}
	\begin{center}
		\includegraphics[width=0.8\textwidth]{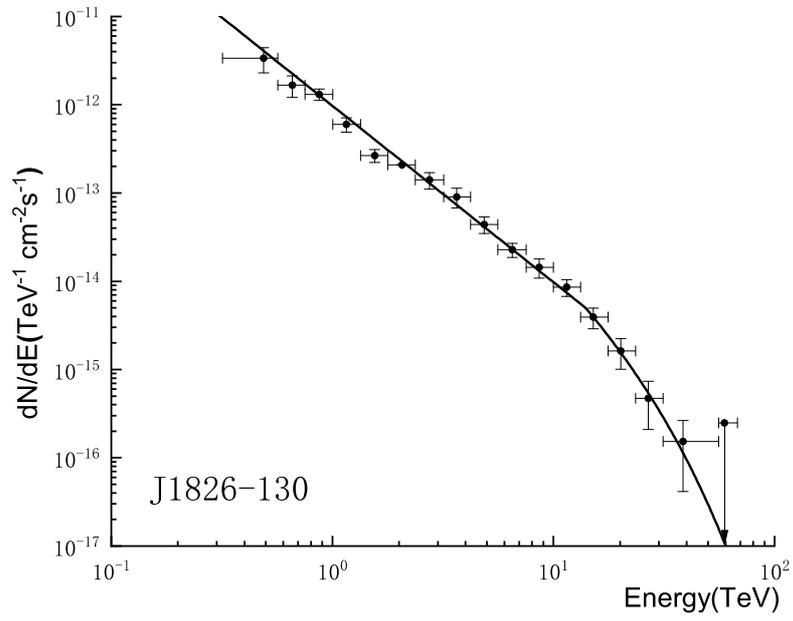} %
		\caption{An incomplete gamma-ray  spectrum of HESS J1826-130. The data are from HESS
			[18]. The parameters of the GC model are $\Gamma_2=2$ and $E_{\pi}^{cut}=18~TeV$ in equation (2.11).
		}\label{fig:9}
	\end{center}
\end{figure}

\begin{figure}
	\begin{center}
		\includegraphics[width=0.8\textwidth]{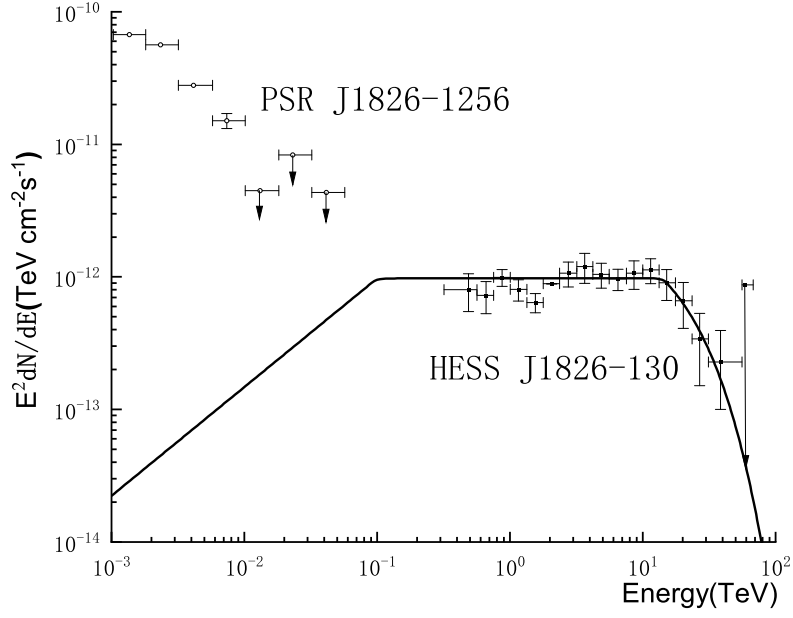} %
		\caption{Gamma-ray spectrum of HESS J1826-130 predicted by the GC model (solid curve).
			The parameter $\beta_p\equiv 0.91$ is fixed by HESS J1640-465.
			The hollow points are related to a neighboring PSR J1826-1256
			[9], which may shadow the spectrum of
			HESS J1826-130 at $E_{\gamma}<0.1~TeV$.
		}\label{fig:10}
	\end{center}
\end{figure}

\begin{figure}
	\begin{center}
		\includegraphics[width=0.8\textwidth]{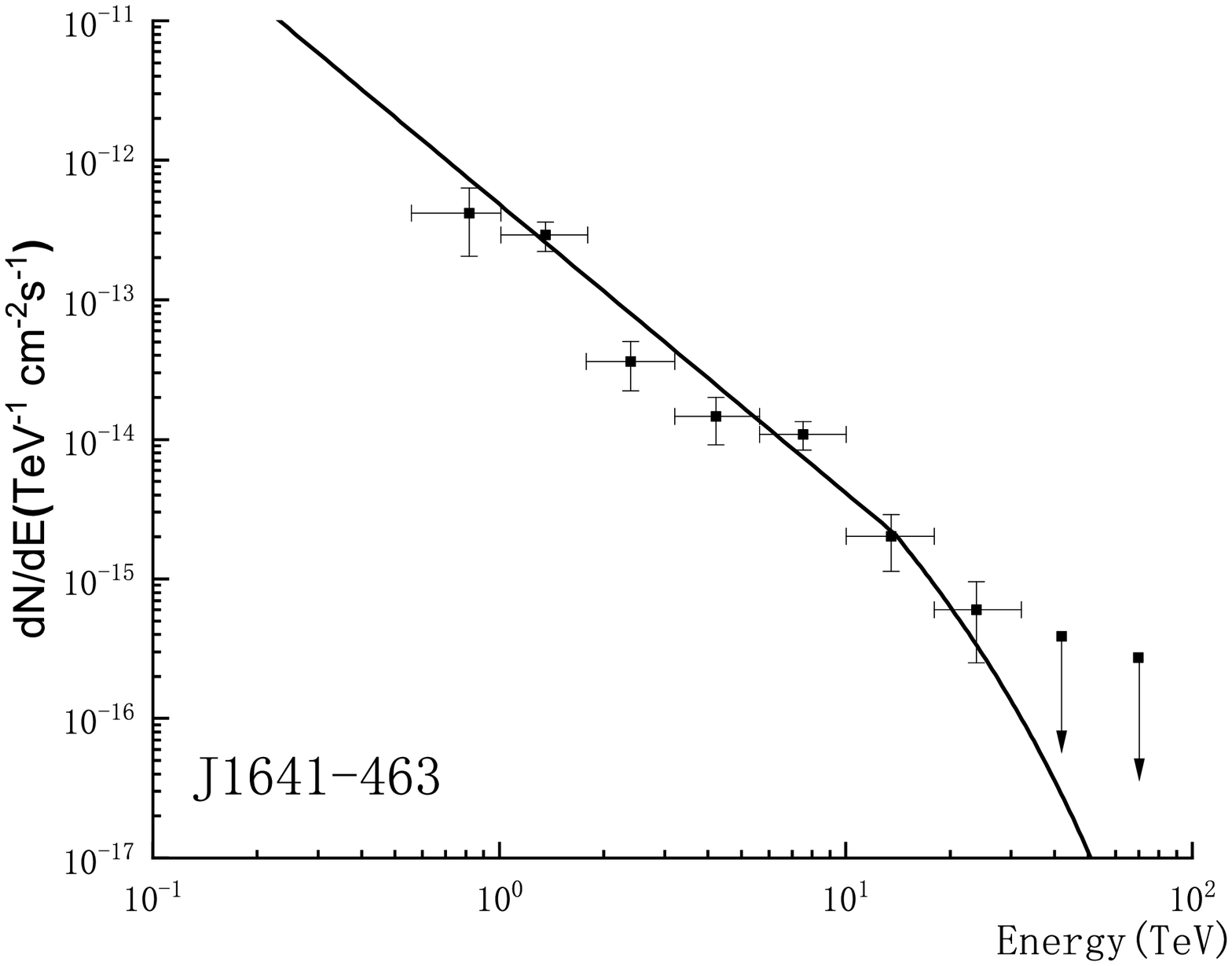} %
		\caption{An incomplete gamma-ray spectrum of HESS J1641-463 [20].
			The parameters of the GC model are $\Gamma_2=2.07$ and $E_{\pi}^{cut}=14~TeV$ in equation (2.11).
		}\label{fig:11}
	\end{center}
\end{figure}

\begin{figure}
	\begin{center}
		\includegraphics[width=0.8\textwidth]{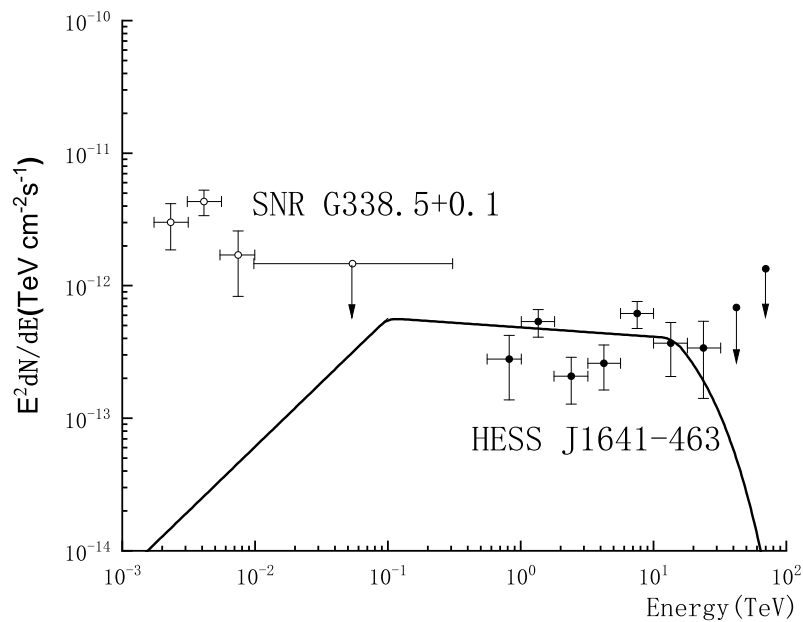} %
		\caption{Gamma-ray spectrum of HESS J1641-463 predicted by the GC model (solid curve).
			The parameter $\beta_p\equiv 1.02$ is fixed by HESS J1640-465.
			The hollow points are related to a neighboring SNR G338.5+0.1 [14], which may shadow the spectrum of
			J1641-463 at $E_{\gamma}<0.1~TeV$.
		}\label{fig:12}
	\end{center}
\end{figure}

\begin{figure}
	\begin{center}
		\includegraphics[width=0.8\textwidth]{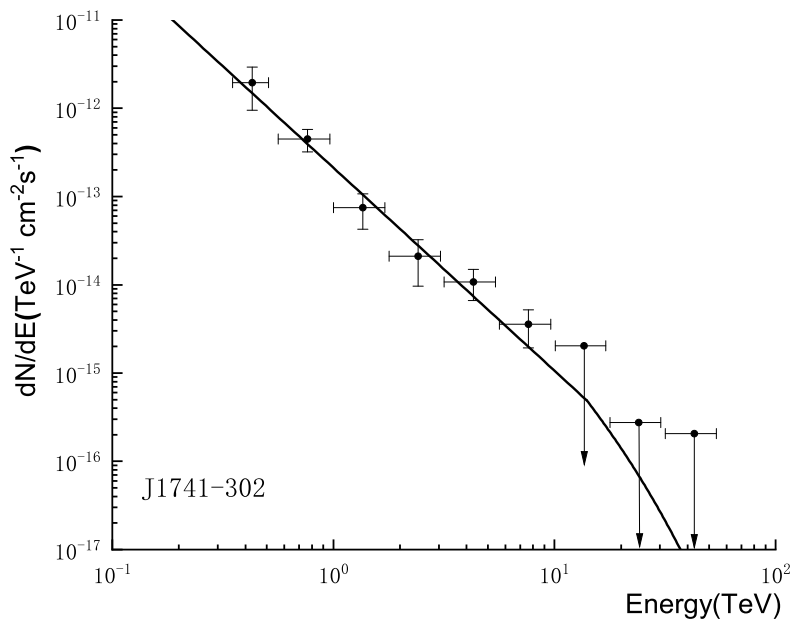} %
		\caption{An incomplete gamma-ray spectrum of HESS J1741-302 [21]. The parameters of the GC model are $\Gamma_2=2.3$ and $E_{\pi}^{cut}=14~TeV$ in equation (2.11).
		}\label{fig:13}
	\end{center}
\end{figure}

\begin{figure}
	\begin{center}
		\includegraphics[width=0.8\textwidth]{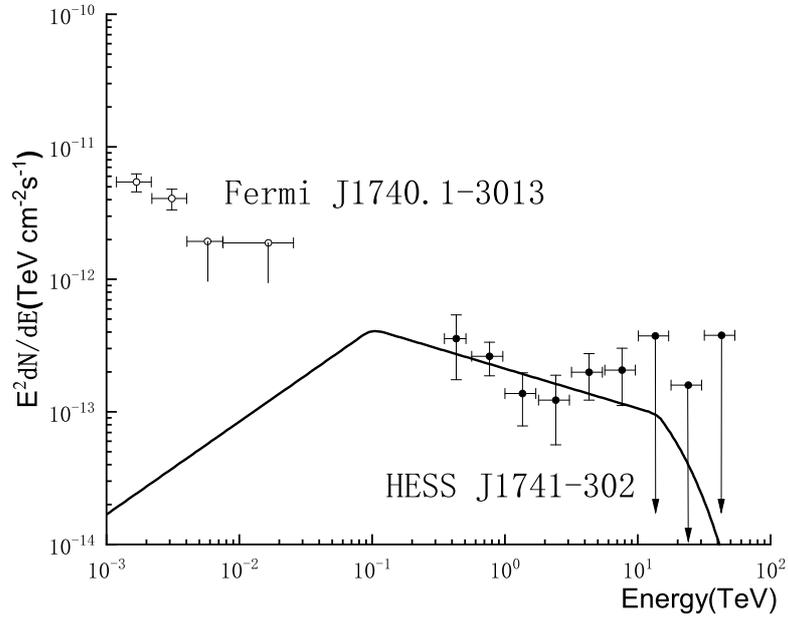} %
		\caption{Gamma-ray spectrum of HESS J1741-302 predicted by the GC model (solid curve).
			The parameter $\beta_p\equiv 1$  refers to HESS J1641-463 and HESS J1826-139.
			The hollow points are  related to a neighboring Fermi J1740.1-3013 [24],
			which may shadow the spectrum of
			J1741-302 at $E_{\gamma}<0.1~TeV$.
		}\label{fig:14}
	\end{center}
\end{figure}

\begin{table}[htbp]
	\caption{Parameters of $\gamma-$ray spectra of pulsars in the GC-model.}\label{tab:1}
	\vskip 0.1cm
	\includegraphics[width=1.05\columnwidth,angle=0]{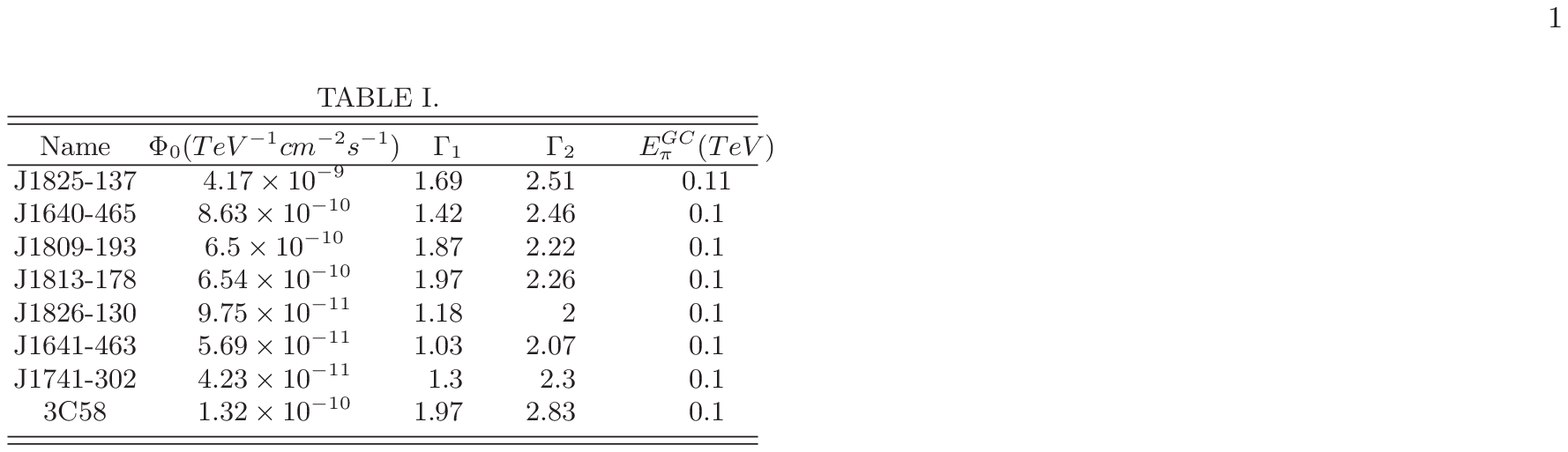}
	
\end{table}

\end{document}